\documentclass[twocolumn,aps,prc,showpacs,superscriptaddress,preprintnumbers,floatfix,nofootinbib]{revtex4}
\usepackage{amsmath,bm}
\usepackage{graphicx}
\begin{document}

\title{Charge separation with fluctuating domains in relativistic heavy-ion collisions}

\author{Qi-Ye Shou}
\affiliation{Shanghai Institute of Applied Physics, Chinese Academy of Sciences, Shanghai 201800, China}
\affiliation{University of Chinese Academy of Sciences, Beijing 100049, China}

\author{Guo-Liang Ma}
\affiliation{Shanghai Institute of Applied Physics, Chinese Academy of Sciences, Shanghai 201800, China}

\author{Yu-Gang Ma}
\affiliation{Shanghai Institute of Applied Physics, Chinese Academy of Sciences, Shanghai 201800, China}


\begin{abstract}

Charge separation induced by the chiral magnetic effect suggested that some ${\cal P}$- or ${\cal CP}$-odd metastable domains could be produced in a QCD vacuum in the early stage of relativistic heavy-ion collisions. Based on a multi-phase transport model, our results suggest that a domain-based scenario with final state interactions can describe the solenoidal tracker at RHIC detector (STAR) measurements of both same- and opposite-charge azimuthal angle correlations, $\left\langle\cos(\phi_{\alpha}+\phi_{\beta})\right\rangle$, in Au+Au collisions at $\sqrt{s_{_{\rm NN}}}=200$ GeV. The occupancy factor of the total volume of domains over the fireball volume is small, which indicates that the size and number of metastable domains should be relatively small in the early stage of a quark-gluon plasma.
 
\end{abstract}


\maketitle

Theoretical studies predicted that some metastable domains may arise from the quark-gluon plasma (QGP) created in relativistic heavy-ion collisions, which could lead to a possible local violation of ${\cal P}$ or ${\cal CP}$~\cite{Lee:1973iz,Lee:1974ma}. These metastable domains represent nontrivial topological solutions of the QCD vacuum~\cite{Kharzeev:2001ev}, where the vacuum transitions are expected to happen via instanton for low temperatures~\cite{'tHooft:1976up,'tHooft:1976fv} or sphaleron for high temperature~\cite{McLerran:1990de}. Consequently, a charge separation of quarks with respect to the reaction plane in non-central heavy-ion collisions can be produced, in the presence of an immensely strong magnetic field with an order of magnitude of $eB \sim m_{\pi}^{2} \sim 10^{18} G$. Such phenomenon is so-called chiral magnetic effect (CME)~\cite{Kharzeev:2007jp}, which has been observed by measuring same- and opposite- charge azimuth correlations, $\left\langle\cos(\phi_{\alpha} \pm \phi_{\beta})\right\rangle$, in experiments at both the BNL Relativistic Heavy Ion Collider (RHIC)~\cite{Abelev:2009ac} and the CERN Large Hadron Collider (LHC)~\cite{Abelev:2012pa}. Its energy dependence is thought as one of the important observables in the study of various aspects of the QCD phase diagram~\cite{Liu:2013eq, Ko:2013mf} in the beam energy scan (BES) program at RHIC~\cite{Mohanty:2011nm}. So far these experimental measurements of $\left\langle\cos(\phi_{\alpha} \pm \phi_{\beta})\right\rangle$ are qualitatively consistent with the CME expectations. Recently, a multiphase transport (AMPT) model was implemented to study charge azimuthal correlations by introducing an initial electric charge dipole distribution, which discloses that final state interactions can play a significant role in weakening the initial charge separation in Au+Au collisions at the top RHIC energy~\cite{Ma:2011uma}.  An $A_{ch}$-dependent charge asymmetry of pion elliptic flow is also reproduced by introducing an initial electric charge quadrupole distribution, which provides a helpful constraint to the quadrupole effect from the chiral magnetic wave~\cite{Ma:2014iva}. Nevertheless, no real dynamical scenario of metastable domains were taken into account in Ref.~\cite{Ma:2011uma}, because the imported dipole charge separation was assumed to be randomly distributed in the global fireball. In this work, we introduce initial separated charges located inside many ${\cal P}$- or ${\cal CP}$-odd bubbles, then some useful information about the properties of metastable domains is extracted by comparing charge azimuth angle correlations between the model and the experimental data.

We use a multi-phase transport model with a string melting scenario to investigate the charge separation with fluctuating ${\cal P}$- or ${\cal CP}$-odd metastable domains. The AMPT model is a dynamical transport model which consists of four main components: initial conditions, partonic cascade, hadronization, and hadronic rescatterings~\cite{Lin:2004en}. The initial conditions, which include the spatial and momentum distribution of participant matter, minijet partons production, and soft string excitations, are obtained from the HIJING model~\cite{Wang:1991hta,Gyulassy:1994ew}. The parton cascade starts the parton evolution with a quark-anti-quark plasma from the melting of strings. Parton scatterings are modelled by Zhang's parton cascade (ZPC), which currently only includes two-body elastic parton scatterings using cross sections from pQCD with screening masses~\cite{Zhang:1997ej}.  A quark coalescence model is used to combine partons into hadrons at freeze-out. The evolution dynamics of the hadronic matter is described by a relativistic transport (ART) model~\cite{Li:1995pra}.  Since the current implementation of the ART model does not conserve the electric charge, only resonance decays are considered and hadronic scatterings are turned off to ensure charge conservation in this study.

\begin{figure}
\center
\includegraphics[scale=0.4]{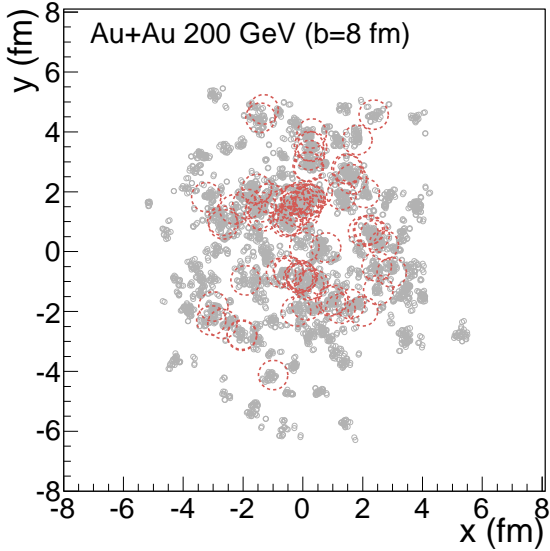}
\caption{(Color online) A schematic illustration of numerous metastable domain bubbles ($r\sim$ 0.5 fm) in an initial partonic fireball for a typical Au+Au event at $\sqrt{s_{_{\rm NN}}}$=200 GeV (b=8 fm), where large dashed circles represent domain bubbles and small solid circles represent partons.}
\label{fig:domains}
\end{figure}

In the previous study by Ma and Zhang~\cite{Ma:2011uma}, in order to separate a percentage of the charges initially, they switch the $p_{y}$ values of a percentage of the downward moving $u$ quarks with those of the upward moving $\bar{u}$ quarks, and likewise for $\bar{d}$ and $d$ quarks. The coordinate system is set up so that the $x$-axis is in the reaction plane and the $y$-axis is perpendicular to the reaction plane with the $z$-axis being the incoming direction of one nucleus, where the reaction plane azimuthal angle $\Psi_{RP}$ is aligned with zero degree. In some sense, the imported separated charges are globally distributed inside the partonic fireball. To improve the global picture of charge separation into a local one of charge separation, a number of domains are intruduced as bubbles  which are placed in the initial quark-anti-quark plasma in this work. These domain bubbles are parameterized by two factors, $r$ and $p_{T}^{eff}$. The former denotes the radii up-limit of a domain bubble while the latter denotes the largest transverse momentum below which partons can be affected by the CME. In this way, we only switch the $p_{y}$ values of two quarks (for which their $p_{T}$ are less than $p_{T}^{eff}$) within the same domain bubble, so that it effectively provides a physical picture of local charge separation. Since the most possible vacuum transitions are those which connect two neighboring vacua (i.e. the winding number $Q_{W}$ = 1) and other transitions between two remote vacua (i.e., $Q_{W} >$ 1) are strongly suppressed, the most likely way of charge separation in each domain is switching $p_{y}$ values of one pair of $u$/$\bar{u}$ and one pair of $d$/$\bar{d}$~\cite{Kharzeev:2007jp}. Namely, four quarks participate in charge separation effect in each domain for $N_{f}$ = 2.  On the other hand, only quarks with transverse momenta $p_{T}^{eff}$~$< 1/r$ are expected to be affected by the CME. Based on these theoretical assumptions, we search through all possible domain bubbles of size $< r$ with at least four proper quarks in the whole partonic fireball and assume that all bubbles are affected by the CME in our simulations. We focus Au+Au collisions at $\sqrt{s_{_\mathrm{NN}}}=200$ GeV with a 10 mb parton interaction cross section. Fig.~\ref{fig:domains} gives a schematic illustration where many metastable domain bubbles ($r\sim$ 0.5 fm) are born in an initial partonic fireball for a typical Au+Au event at $\sqrt{s_{_{\rm NN}}}$=200 GeV (b=8 fm).

\begin{figure}
\center
\includegraphics[scale=0.45]{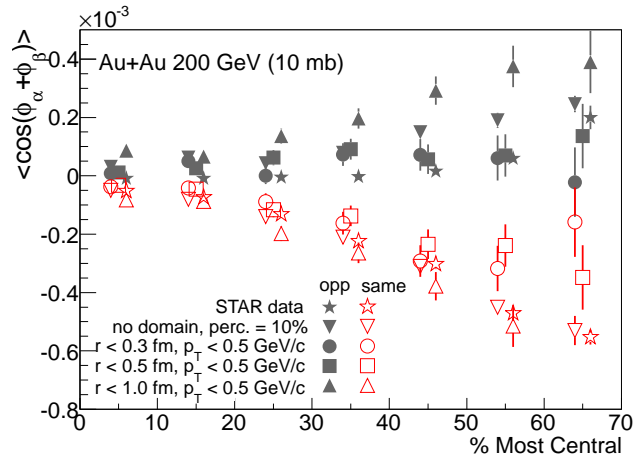}
\caption{(Color online) Centrality dependence of $\left\langle\cos(\phi_{\alpha}+\phi_{\beta})\right\rangle$ in Au+Au collisions at $\sqrt{s_{_{\rm NN}}}$=200 GeV (with a 10 mb parton interaction cross section). The different symbols represent the results from different settings of AMPT models, and the stars represent experimental data~\cite{Abelev:2009ac,Abelev:2009ad}. Some points are slightly shifted for clarity.
}
\label{fig:rSum}
\end{figure}

It has been proposed by Voloshin that charge separation likely coming from local parity violation could be measured via charge azimuthal correlation $\left\langle\cos(\phi_{\alpha}+\phi_{\beta}-2\Psi_{RP})\right\rangle$, where $\phi_{\alpha}$ and $\phi_{\beta}$ are the emission azimuthal angles of charged particles $\alpha$ and $\beta$, the subscripts of $\alpha$ and $\beta$ denote the signs of electric charges which can be either positive or negative~\cite{Voloshin:2004vk}. Observations from the RHIC-STAR experiment are manifested as the consistency with the expectation of the CME~\cite{Abelev:2009ac,Abelev:2009ad}. Phenomenologically, the previous work performed by Ma and Zhang found that the initial charge separation can be significantly reduced by the final interactions such as parton cascade, hadronization, and resonance decay~\cite{Ma:2011uma}. In the basis of the previous study~\cite{Ma:2011uma}, our following calculations take both final state interactions and a dynamic scenario with fluctuating domains into account.

In Fig.~\ref{fig:rSum}, the charge azimuthal correlation of $\left\langle\cos(\phi_{\alpha}+\phi_{\beta})\right\rangle$ as a function of centrality bin from the AMPT simulations is presented, where open symbols represent same-charge correlations and solid symbols represent opposite-charge correlations, and the STAR experiment data are denoted by stars. Ma and Zhang's previous calculations which implemented a percentage of 10\% for initial charge separation, are also shown by down-triangles. For same-charge correlations, the magnitudes of azimuthal correlations look similar between $r < $ 0.3 fm and $r < $ 0.5 fm, and a little larger for $r < $ 1.0 fm, and all of them can describe the data. For opposite-charge correlations, we found that the simulation results with $r <$ 0.3 or 0.5 fm and $p_{T}^{eff}<$ 0.5 GeV/c can depict the data much better than those with $r <$ 1 fm.  On the other hand, the results with fluctuating domains of $r <$ 0.3 or 0.5 fm provide a closer description to the opposite-charge data than the previous calculations without domains which overestimate the opposite-charge data especially for more peripheral collisions. 

\begin{figure}
\center
\includegraphics[scale=0.45]{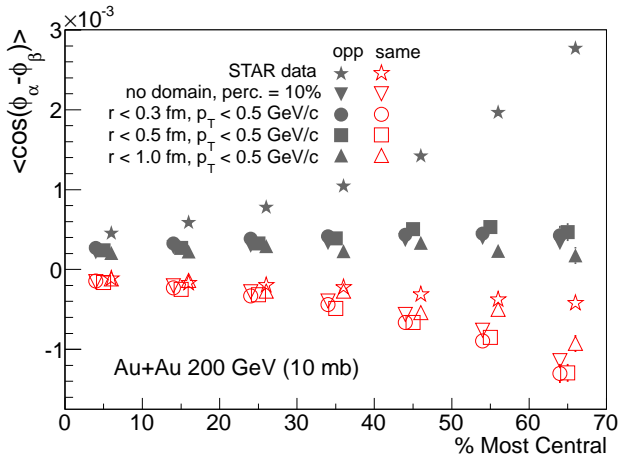}
\caption{
(Color online) Same as Fig.~\ref{fig:rSum}, but for $\left\langle\cos(\phi_{\alpha}-\phi_{\beta})\right\rangle$.
}
\label{fig:rDiff}
\end{figure}

Another reaction plane independent correlation, $\left\langle\cos(\phi_{\alpha}-\phi_{\beta})\right\rangle$, is shown in Fig.~\ref{fig:rDiff}. The STAR measurement shows that the opposite-charge correlation increases and the same-charge correlation decreases from central to peripheral collisions. The previous AMPT work without domains shows the same trends, but is not able to match the experimental data in magnitude. Our calculations show there are small differences between different settings of domains, and all of them still have much lower values than experiment measurements. It reveals that the domain effect is not the key to compensate the large difference, and the contribution from non-negligible backgrounds needs further investigation for the reaction plane independent correlations~\cite{Ma:2011uma}.

\begin{figure}
\center
\includegraphics[scale=0.45]{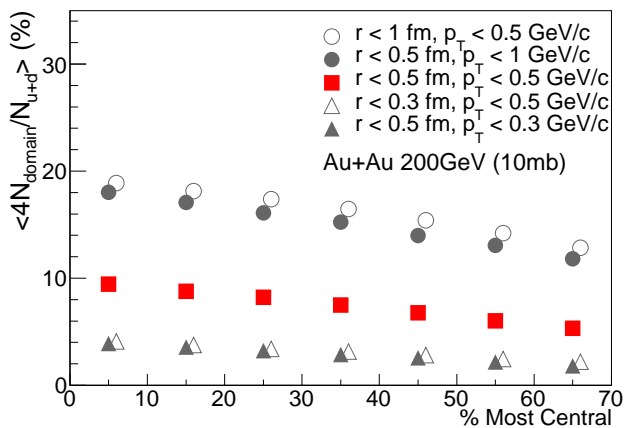}
\caption{(Color online) Centrality dependences of production rates of different settings of domains in Au+Au collisions at $\sqrt{s_{_{\rm NN}}}$=200 GeV.}
\label{fig:domainRate}
\end{figure}

To compare with Ma and Zhang's work~\cite{Ma:2011uma} in which a percentage is employed to describe the strength of initial global charge separation,  a domain production rate is defined as $\left\langle 4N_{domain}/N_{u+d} \right\rangle$, where {$N_{domain}$} and {$N_{u+d}$} denote the number of domains and the total number of $u+d$ and their anti-quarks $\bar{u}+\bar{d}$, respectively, and $<>$ is event averaged. In this sense, the domain rate corresponds to the initial charge separation percentage in the previous work. Fig.~\ref{fig:domainRate} presents the domain rates as functions of centrality bin for different $r$ and $p_{T}^{eff}$, if all possible bubbles are affected the CME. It is obvious to see that the domain rate decreases as the centrality goes from central to peripheral. For the case of $r < $ 0.5 fm and $p_{T}^{eff} <$ 0.5 GeV/c, the domain rate decreases from 10\%(most central) to 7\%(most peripheral), which is basically consistent with the previous global charge separation percentage of 10\%. It is certainly possible to further improve the description by adjusting the centrality dependence of the domain rate. However, we would like to emphasize that our goal was not to fit precisely the data. Our objective was to check if a domain-based charge separation, with the final interactions, can better reproduce the experimental data than the global charge separation.

In addition, it is also significant to study an occupancy factor of the total volume of domains over the whole volume of partonic fireball. Tab.~\ref{tab:domainV} provides the occupancy factors for different centrality bins and domain radii, where each domain is approximately considered as a sphere and the volume of the fireball is extracted as an ellipsoid from initial parton coordinate space. The data-matching case of $r <$ 0.5 fm shows the factors of 2\%-4\%, while off-matching case of $r <$ 1 fm reaches about 40\%-70\%. It indicates that the size and number of metastable domains should be relatively small in the early stage of QGP. 

\begin{center}
\begin{table}[h]
\caption{The occupancy factors of the total volume of domains over the fireball volume as functions of the centrality bin in Au+Au collisions at $\sqrt{s_{_{\rm NN}}}$ = 200 GeV.}

\centering
\begin{tabular}{@{}*{7}{l}}
\hline
\% most Central & $r <$ 0.3 fm & $r <$ 0.5 fm &$ r<$ 1 fm\\
\hline
0-10&0.38 \%&4.15 \%&65.81 \%\\
10-20&0.40 \%&4.29 \%&70.62 \%\\
20-30&0.36 \%&4.00 \%&67.31 \%\\
30-40&0.31 \%&3.35 \%&59.71 \%\\
40-50&0.26 \%&2.69 \%&48.79 \%\\
50-60&0.19 \%&2.20 \%&42.33 \%\\
60-70&0.21 \%&2.19 \%&41.66 \%\\
\hline
\label{tab:domainV}
\end{tabular}
\end{table}
\end{center}

It should be noticed that all partons are valence quarks from the string melting mechanism in the AMPT model. Although it is unphysical since gluons dominate the initial stage of relativistic heavy-ion collisions, our results should depend more on the effect of partonic scatterings than on the composition of the partonic matter, because gluons do not carry charge and thus are not affected by the chiral magnetic effect. Therefore, we expect that our results persist if the quarks outside the domains are replaced by gluons to imitate a realistic picture of fluctuating domains in a gluon-rich partonic matter.

In summary, we have investigated the charge separation with fluctuating ${\cal P}$- or ${\cal CP}$-odd metastable domains based on a multi-phase transport model in Au+Au collisions at $\sqrt{s_{_{\rm NN}}}=200$ GeV. It is found that the experimental measurements of the azimuthal angle correlation $\left\langle\cos(\phi_{\alpha}+\phi_{\beta})\right\rangle$ can be well described when only light quarks with $p_\mathrm{T}<$ 0.5 GeV/c are affected within many metastable domains of $r<$ 0.3 or 0.5 fm. Our results demonstrate that besides final state interactions which play a role in reducing the charge separation, a domain-based scenario is important in describing the data for both same- and opposite- charge correlations, owing to a centrality-dependent domain production rate. However, the significant difference for the correlation $\left\langle\cos(\phi_{\alpha}-\phi_{\beta})\right\rangle$ between experimental data and the model calculation needs to be further investigated. Our simulation results also show that the occupancy factor of the total volume of all domains over the whole volume of a partonic fireball is small, which indicates that the size and number of metastable domains should be relatively small in the early stage of QGP. However, it is worth noticing that the metastable domains are introduced by hand and the partonic system evolves without considering the electromagnetic field in our approach. It would be interesting to establish our model by introducing a real initial domain distribution from the CME and developing the QGP within a chiral kinetic theory~\cite{Chen:2012ca,Son:2012zy} for future investigations.

\hspace{-5mm}

We thank Che-Ming Ko and Bin Zhang for helpful discussions, and the HIRG PC farm for computer time. This work was supported by the Major State Basic Research Development Program in China under Contract No. 2014CB845400, the NSFC under Projects No. 11375251, No. 11220101005, No. 11175232, and No. 11035009, the Knowledge Innovation Program of CAS under Grant No. KJCX2-EW-N01, CCNU-QLPL Innovation Fund under Grant No. QLPL2011P01, and the ``Shanghai Pujiang Program" under Grant No. 13PJ1410600.

{}
\end{document}